\documentclass[aps,twocolumn,epsf,floats,prl]{revtex4}
\usepackage{graphics,graphicx,epsfig}
\usepackage{amssymb}
\usepackage{epsf,epstopdf,wrapfig}

\newcommand{\sgn}{\mbox{sgn}}

\begin{document}

\title{Dimensionality and  dynamics in the behavior of {\em C. elegans}}

\author{Greg J Stephens,$^{a-c}$ Bethany Johnson--Kerner,$^{a}$ William Bialek$^{a,b}$ and William S Ryu$^{a}$}

\affiliation{$^a$Lewis--Sigler Institute for Integrative Genomics,
$^b$Joseph Henry Laboratories of Physics, and
$^c$Center for the Study of Brain, Mind and Behavior,Princeton University,
Princeton, New Jersey 08544 USA}

\date{\today}

\begin{abstract} 
A major challenge in analyzing animal behavior is to discover some underlying simplicity in complex motor actions.  Here we show that the space of shapes adopted by the nematode
 {\em C. elegans}  is surprisingly low dimensional, with just four 
dimensions accounting for $95\%$ of the shape variance, and we partially reconstruct `equations of motion' for the dynamics in this space.
These dynamics have multiple attractors, and we find that the worm visits these in a rapid and almost completely deterministic response to weak thermal stimuli.  Stimulus-dependent correlations among the different modes suggest that one can generate more reliable behaviors by synchronizing stimuli to the state of the worm in shape space.  We confirm this prediction, effectively ``steering'' the worm in real time.
\end{abstract}

\maketitle

\section{Introduction}

The study of animal behavior is rooted in two divergent traditions.  One approach creates well-controlled situations, in which animals are forced to choose among a small discrete set of behaviors, as in psychophysical experiments \cite{green+swets_66}.  The other, taken by ethologists \cite{ethology},  describes the richness of the behaviors seen in more natural contexts.  One might hope that simpler organisms provide model systems in which the tension between these approaches can be resolved, leading to a fully quantitative description of complex, naturalistic behavior.

Here we explore the motor behavior of the nematode, {\em Caenorhabditis elegans}, moving freely on an agar plate \cite{geng+al_03,cronin+al_05,hoshi+shingai_06,huang+al_06,karbowski+al_06}.  Though lacking the full richness of a natural environment,  this unconstrained motion allows for complex patterns of spontaneous motor behaviors \cite{croll_75}, which are modulated in response to chemical, thermal and mechanical stimuli \cite{bargmann_93,hedgecock+russell_75,chalfie+al_85}.  
Using video microscopy of the worm's movements, we find a low dimensional but essentially complete description of the macroscopic motor behavior.
Within this low dimensional space we reconstruct equations of motion which reveal multiple attractors---candidates for a rigorous definition of behavioral states.  
We show that these states are visited as part of a surprisingly reproducible response of {\em C. elegans} to small temperature changes.  Correlations among fluctuations along the different behavioral dimensions suggest that some of the randomness in the behavioral responses could be removed if sensory stimuli are delivered only when the worm is at a well defined initial state. We present experimental evidence in favor of this idea, showing that worms can be ``steered'' in real time by appropriately synchronized stimuli.

\section{Eigenworms}
\begin{figure}[b]
\begin{center}
\includegraphics[width=0.85\columnwidth,keepaspectratio=true]{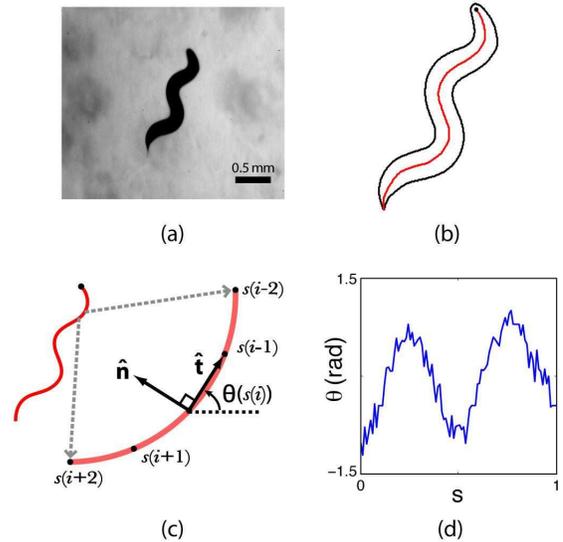}
\end{center}
\caption{ Describing the shapes of worms. (a) Raw image
 in the tracking microscope. (b) We process
the image in (a) and pass a curve through the center of the
body.  The black circle marks the head. (c) We measure distance along the curve (arclength $s$) in
normalized units, and at each point we define the tangent
${\bf\hat{t}}(s)$ and and normal ${\bf\hat{n}}(s)$ to this curve. The tangent points in
a direction $\theta(s)$, and variations in this angle correspond to the
curvature $\kappa(s) = d\theta(s)/ds$. (d) We rotate all images so that
the average angle $\theta$ is zero; therefore $\theta(s)$ provides a description
of the wormÕs shape that is independent of our coordinate system, and intrinsic to the worm itself. }
\label{fig:shapes}
\end{figure}

We use tracking microscopy with high spatial and temporal resolution to extract the two-dimensional shape of individual {\em C. elegans} from images of
freely moving worms over long periods of time (Fig 1a; see Methods).  Variations in the thickness of the worm are small, so we describe the shape by a curve that passes through the center of the body (Fig 1b).  We measure position along this curve (arc length) by the variable $s$, normalized so that $s=0$ is the head and $s=1$ is the tail.  The position of the body element at $s$ is denoted by ${\bf x}(s)$, and we sample this function at $N=100$ equally spaced points along the body.  These variables provide an essentially complete description of the motor output.

We analyze the worm's shapes in a way intrinsic to its own behavior, not to our arbitrary choice of coordinates.  The intrinsic geometry of a curve in the plane is defined by the Frenet equations \cite{frenet,struik_61},
\begin{eqnarray}
{{d{\bf x}(s)}\over{ds}} &=& {\bf \hat t}(s) ,\\
{{d{\bf \hat t}(s)}\over{ds}} &=& \kappa (s) {\bf \hat n}(s) ,
\end{eqnarray}
where ${\bf t}(s)$ is the unit tangent vector to the curve, ${\bf \hat n}(s)$ is the unit normal to the curve, and $\kappa (s)$ is the scalar curvature (Fig 1b).  If the tangent vector points in a direction $\theta (s)$, then $\kappa (s) = d\theta(s)/ds$.  Curvature as a function of arc length, $\kappa (s)$, thus provides a ``worm--centered'' description, but in practice this involves taking two derivatives and thus is noisy.  As an alternative, we describe the curve by   $\theta (s)$, but remove the dependence on our choice of coordinates by rotating each image so that the mean value of $\theta$ along the body always is zero (Fig 1c); this rotated version of $\theta (s)$ contains exactly the same information as $\kappa(s)$.

Although the worm has  no discrete joints, we expect that the combination of elasticity in the worm's  body wall and a limited number of muscles will lead to a limited effective dimensionality of the shape and motion.   In the simplest case, the relevant low dimensional space will be a Euclidean projection of the original high dimensional space.  If this is true, then the covariance matrix of angles, $C(s, s' ) = \langle(\theta(s)-\langle\theta\rangle)( \theta (s')-\langle\theta\rangle)\rangle$ will have only a small number of nonzero eigenvalues.    Figure 2a shows the covariance matrix, and its smooth structure is a strong hint that there will be only a small number of significant eigenvalues; this is shown explicitly in Fig 2b.  Quantitatively, if we measure the total variance in angle along the body,  then over $95\%$ of this variance can be accounted for by just four eigenvalues.

\begin{figure}
\begin{center}
\includegraphics[width=0.9\columnwidth,keepaspectratio=true]{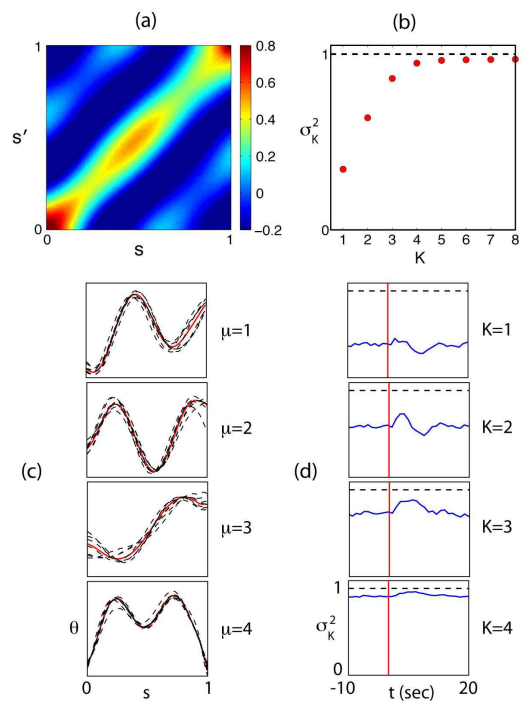}
\end{center}
\caption{ Covariance of shape fluctuations and eigenworms. (a) The covariance matrix of fluctuations in angle $C(s, s')$
is computed from $60,000$ images across eight different worms. The inhomogeneity along the diagonal shows that the Ònormal modesÓ  of the motion are not sinusoidal but the
smooth structure of $C(s,s')$ means that a small number of modes are significant. (b) We find the eigenvalues
of $C(s,s')$ and compute $\sigma^2_K$, the fraction of the total variance  captured by keeping just $K$
modes (see Methods); $95\%$ of the variance is accounted for by the first four modes. (c) Associated with each of the dominant
modes is an eigenvector describing angle vs.\ position along the worm; we refer to these as the eigenworms $u_\mu(s)$. The population-mean eigenworms (red)
are highly reproducible across individual worms (black). (d) In response to strong thermal stimuli the worm's shape remains 
within the four dimensional subspace identified from analysis of spontaneous crawling.  Worm images are recorded at times 
synchronized to a heat pulse (see Methods) and we display $\sigma_K^2$ aligned with
the stimulus.  Though the shapes vary systematically in response to these stimuli, the accuracy of our reconstructions does not  and using 
all four modes we continue to account for $95\%$ of the shape variance.}
\label{fig:modes}
\end{figure}

Associated with each of the eigenvalues $\lambda_\mu$ is an eigenvector $u_\mu (s)$, sometimes referred to as a `principal component' of the function $ \theta (s)$.  If only $K=4$ eigenvalues are significant,
then we can write the shape of the worm as a superposition of `eigenworm' shapes,
\begin{equation}
\theta (s) \approx \sum_{\mu =1}^K a_\mu u_\mu (s) ,
\label{recon}
\end{equation}
where the four variables $\{a_\mu\}$ are the amplitudes of motion along the different principal components, $a_\mu = \sum_s u_\mu(s) \theta (s)$.  We see in Fig 2c that these modes are highly reproducible from individual to individual.

\begin{figure}[thb]
\begin{center}
\includegraphics[width=.95\columnwidth,keepaspectratio=true]{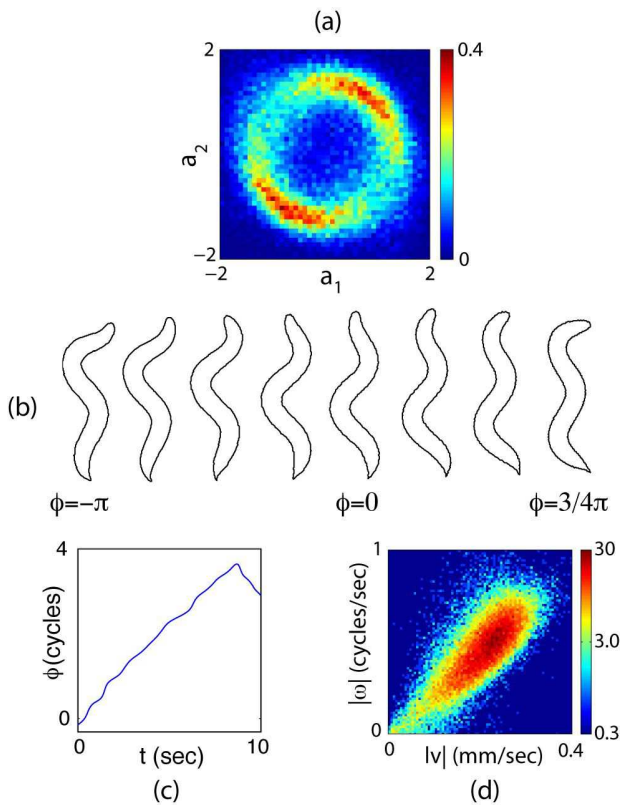} 
\end{center}
 \caption{ Motions along the first two eigenworms. (a)  The joint probability
density of the first two amplitudes, $\rho(a_1, a_2)$, with
units such that $\langle a_1^2 \rangle=\langle a_2^2\rangle=1$.
The ring structure suggests that these modes form an
oscillator with approximately fixed amplitude and varying
phase $\phi=\tan^{-1}{\left(-a_2/ a_1\right)}$.  (b) Images of worms with different values of $\phi$ show that variation in
phase corresponds to propagating a wave of bending
along the wormÕs body. (c) Dynamics of the phase $\phi(t)$
shows long periods of linear growth, corresponding
to a steady rotation in the $\{a_1, a_2\}$ plane, with occasional, abrupt reversals.  (d) The joint density $\rho(|v|,|\omega|)$.  The phase velocity
$ \omega= d\phi/dt$ in shape space predicts the speed at which the worm crawls on the agar.  The crawling speed was defined as the time derivative 
of the worm's center of mass.}
\label{fig:a1a2}
\end{figure}

Thus far we have considered only worms moving in the absence of deliberate sensory stimuli.  Do the worms continue to move in just a four dimensional shape space when they respond to strong inputs?  To test this, we delivered intense pulses of heat (see Methods), which are known to trigger escape responses \cite{pain}.  We see in Fig 2d that we still account for $\sim95\%$ of the shape variance using just four modes, even though the distribution of shapes during the thermal response is very different from that seen in spontaneous crawling.  We conclude that our four eigenworms provide an effective, low dimensional coordinate system within which to describe {\em C. elegans} motor behavior.

\section{What do the modes mean?}

The projection of worm shapes onto the low-dimensional space of eigenworms provides a new and quantitative foundation for the classical, qualitative descriptions of {\em C. elegans} behavior \cite{croll_75}.  The first two modes are sinuous (although not exactly sinusoidal) oscillations of the body shape (Fig 2c); they form a quadrature pair, so that different mixtures of the two modes correspond to different phases of a wave along the body.  Indeed, the probability distribution of the mode amplitudes, 
$\rho(a_1 , a_2)$, shows a ring of nearly constant amplitude (Fig 3a).   Sampling images around this ring reveals a traveling wave along the body (Fig 3b).  There are relatively long periods of time where the shape changes by a continuous accumulation of the phase angle (Fig 3c), and the speed of this rotation predicts the speed at which the worm crawls (Fig 3d).  

In contrast to the first two modes, the third mode $u_3(s)$ contributes to a nearly constant curvature throughout the middle half of the body (Fig 2c).  The distribution of the mode amplitude $a_3$ has a long tail (Fig 4a), and body shapes chosen from these tails (Fig 4b) exhibit the $\Omega$ configuration classically identified with turning behavior \cite{croll_75}.  Large amplitudes of $a_3$ also correspond to regions of high curvature in the worm trajectory along the agar (Fig 4c).

\begin{figure}[b]
\begin{center}
\includegraphics[width=0.9\columnwidth,keepaspectratio=true]{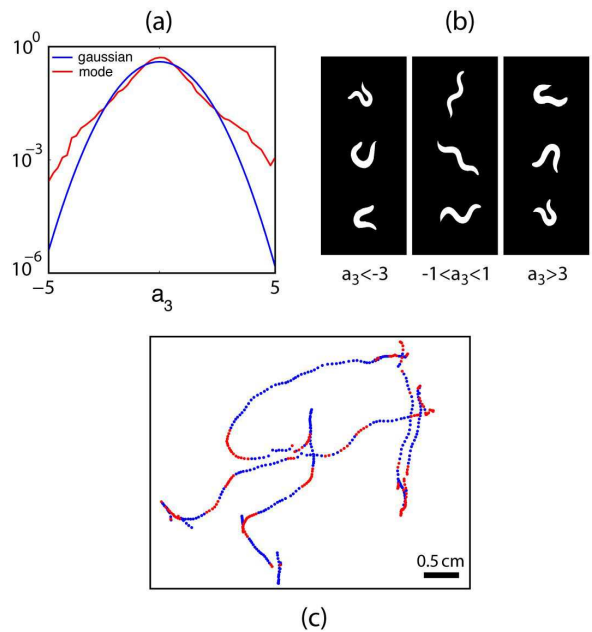}
\end{center}
\caption{ Motions along the third eigenworm. (a) The distribution
of amplitudes $\rho(a_3)$, shown on a logarithmic scale.
Units are such that $\langle a^2_3 \rangle=1$, and
for comparison we show the Gaussian distribution; note the
longer tails in $\rho(a_3)$. (b) Images of worms with values of $a_3$
in the negative tail (left), the middle (center) and positive tail (right). Large
amplitudes of $a_3$ correspond to $\Omega$-like shapes. (c) A two minute
 trajectory of the center of mass sampled at 4\,Hz. Periods where $|a_3| > 1$ are colored red, 
illustrating the association between turning and large displacements along this mode.}
\label{fig:a3}
\end{figure}
The connections between mode amplitudes and the motion of the worm along the agar---as in Figs \ref{fig:a1a2}d and \ref{fig:a3}c---are genuine tests of the functional meaning of our low dimensional description. Quite explicitly, our analysis of worm shapes is independent of the extrinsic coordinates and hence our definition of modes and amplitudes is blind to the actual position and orientation of the worm.  Of course, in order to move the worm must change shape, and our description of the shape in terms of mode amplitudes captures this connection to movement.  Thus, to crawl smoothly forward or backward the worm changes its shape by rotating clockwise or counterclockwise in the plane formed by the mode amplitudes $a_1$ and $a_2$; the speed of crawling is set by the speed of the rotation.  Similarly, to change direction the worm changes shape toward larger magnitudes of the mode amplitude $a_3$, and we see this connection even without defining discrete turning events.

\section{Attractors and behavioral states}

The eigenworms provide a coordinate system for the postures adopted by {\em C. elegans} as it moves; to describe the dynamics of movement we need to find equations of motion in this low dimensional space.  We start by focusing on the plane formed by the first two mode amplitudes $a_1$ and $a_2$. Figure \ref{fig:a1a2} suggests that within this plane the system stays 
at nearly constant values of the radius, so that the relevant dynamics involves just the phase angle $\phi (t)$.  To account for unobserved and random influences these equations need to be stochastic, and to support both forward and backward motion they need to form a system of at least second order.  Such a system of equations would be analogous to the description of Brownian motion using the Langevin equation \cite{stat_mech_text,flyvbjerg}.  Thus we search for equations of the form
\begin{eqnarray}
{{d\phi(t)}\over{dt}} &=& \omega (t), \nonumber \\
{{d\omega(t)}\over{dt}} &=& F[ \phi(t),\omega(t)] + \sigma[ \phi(t),\omega(t)] \eta (t) .
\label{dyn2}
\end{eqnarray}
Here  $F[ \phi(t),\omega(t)] $ defines the average acceleration as a function of the phase and phase velocity, by analogy to the force on a Brownian particle.  The noise is characterized by a random function $\eta (t)$ which we hope will have a short correlation time, and we allow the strength of the noise $\sigma[ \phi(t),\omega(t)] $ to depend on the state of the system, by analogy to a temperature that depends on the position of the Brownian particle.

\begin{figure}[t]
\begin{center}
\includegraphics[width=0.95\columnwidth,keepaspectratio=true]{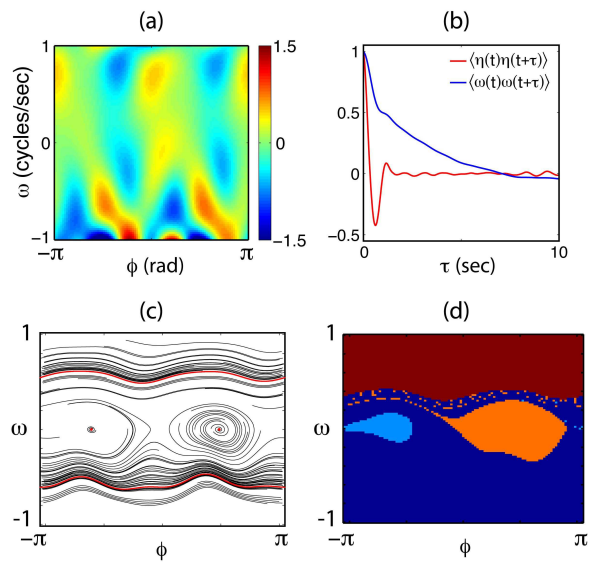} 
\end{center}
\caption{Reconstructing the phase dynamics. (a) The mean
acceleration of the phase as a function of the instantaneous
phase and phase velocity, $F(\omega, \phi)$ in Eq (5). (b)
The correlation function of the noise in the phase dynamics,
$\langle \eta(t) \eta(t+\tau) \rangle$ with $\eta(t)$ defined from Eq (5). The
correlations are confined to very short times, especially when
compared with the correlations of the phase velocity itself.
(c) Trajectories in the deterministic dynamics.  A selection of early-time trajectories are shown in black.  At late times these same trajectories collapse to one of four attractors (red): forward and backward crawling and two pause states. (d) Attracting basins.  Initial conditions covering the $\{\phi,\omega\}$ plane were evolved to long times and each initial phase point was colored corresponding to its asymptotic attractor.  Note that the basins are asymmetric so that some forward motions fall into the reversal or pause attractor.  This highlights the difference between a dynamical definition of behavioral state and phenomenological one based on an instantaneous value such as  $\sgn(\omega(t))$. }
\label{fig:dynamics}
\end{figure}

In Figure \ref{fig:dynamics}a we show our best estimate of the mean acceleration $F[\phi , \omega]$ (see Methods for details).  Once we know $F$, we subtract this mean acceleration from the instantaneous acceleration to recover trajectories of the noise, and the correlation function of this noise is shown in Fig 5b.  The correlation time of the noise is short, which means that we have successfully separated the dynamics into two parts: a deterministic part, described by the function $F[\phi,\omega]$, which captures the average motion in the $\{a_1,a_2\}$ plane and hence the relatively long periods of constant oscillation, and a rapidly fluctuating part $\eta(t)$ that describes ``jittering'' around this simple oscillation as well as the random forces that lead to jumps from one mode of motion to another.

We can imagine a hypothetical worm which has the same deterministic dynamics as we have found for real worms, but no noise.  We can start such a noiseless worm at any combination of phase and phase velocity, and follow the dynamics predicted by Eqs (\ref{dyn2}), but with $\sigma=0$.  The results are shown in Fig \ref{fig:dynamics}c.  The dynamics are diverse on short time scales, depending in detail on the initial conditions, but
eventually all initial conditions lead to one of a small number of possibilities (Fig \ref{fig:dynamics}d):  either the phase velocity is always positive, always negative, or decays to zero as the system pauses at one of two stationary phases.  Thus, underneath the continuous, stochastic dynamics we find four discrete attractors which correspond to well defined classes of behavior.  

\section{Pause states and reproducibility}

The behavior of {\em C. elegans}, particularly in response to sensory stimuli, traditionally has been characterized in probabilistic terms:  worms respond by changing the probability of turning or reversing \cite{Lockery03,Ryu02,gray+al_05}.  This randomness could reflect an active strategy on the part of the organism, or it could reflect the inability of the nervous system to distinguish reliably between genuine sensory inputs and the inevitable background of noise.  Our ability to describe motor behavior more completely, with high time resolution, offers us the opportunity to revisit the ``psychophysics'' of {\em C. elegans}.

\begin{figure}[h]
\begin{center}
\includegraphics[width=.95\columnwidth,keepaspectratio=true]{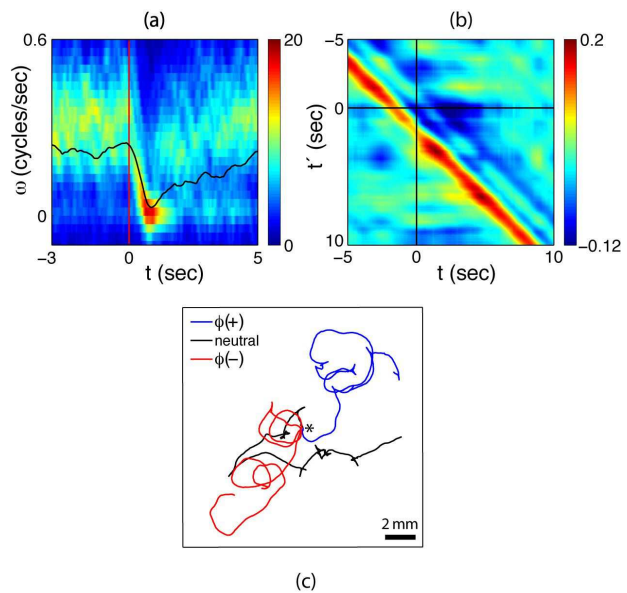} 
\end{center}
\caption{Thermal responses, mode coupling and active steering (see Methods for details).
(a) The distribution of phase velocities $\rho_t(\omega)$ in response to
a brief thermal stimulus.  These data are taken from a collection of 13 worms, each
stimulated with 20 repetitions of a $\Delta T = 0.1^\circ \rm{C}$ pulse. Note that
within one second, the distribution becomes highly concentrated
near $\omega=0$, corresponding to the pause states identified in Fig 5. In this experiment only, the worms were raised at a lower temperature leading to a lower average $\omega$ before the thermal stimulus.  (b) Correlations between phase in the $\{a_1, a_2 \}$ plane
and $a_3$, $C(t,t')=\langle \delta a_3(t) \delta \sin(\phi(t')) \rangle$.  Shortly after the thermal impulse, the modes develop 
a strong anti-correlation which is distinct from normal crawling (c) Worm `steering'.  A thermal impulse conditioned on the instantaneous phase was delivered automatically and repeatedly, causing an orientation change $\dot\Theta$ in the worm's trajectory.  In this example lasting 4 minutes, asynchronous impulses produced a time-averaged 
orientation change $\langle \dot\Theta \rangle= 0.01$\,rad/s (black), impulses at positive phase produced a trajectory with $\langle\dot\Theta\rangle=0.10$\,rad/s (blue), and impulses at negative phase produced $\langle\dot\Theta \rangle=-0.12$\,rad/s (red). This trajectory response is consistent with the early-time mode correlations seen in Fig 6b. We found 13 out of 20 worms produced statistically different orientation changes under stimulated and non-simulated conditions while 
only 1 out of 20 worms responded in the same fashion when the phase was randomized. The asterisk marks the start of the trajectories.} 
\label{fig:responses}
\end{figure}

We consider the response to brief (75 ms), small ($\Delta T \sim 0.1^{\circ}\rm{C}$) changes in temperature, induced by pulses from an infrared laser (see Methods).  These stimuli are large enough to elicit responses  \cite{hedgecock+russell_75} but well below the threshold for pain avoidance \cite{pain}.   In Figure \ref{fig:responses} we show the distribution $\rho_t(\omega)$ of phase velocities as a function of time relative to the thermal pulse.  All of the worms were crawling forward at the moment of stimulation, so the initial phase velocities are distributed  over a wide range of positive values.  Within one second, the distribution narrows dramatically, concentrating near zero phase velocity-the pause states described above.

Arrival in the pause state is  stereotyped both across trials and across worms.
By analogy with conventional psychophysical methods \cite{green+swets_66}, we can ask how reliably an observer could infer the presence of the heat pulse using the worm's response. We find  that just measuring the phase velocity $\omega$ at single moment in time after the pulse is sufficient to provide $\sim 75\%$ correct detection of this small temperature change in single trials.

\section{Coupling the modes and steering the worm}

Our discussion thus far has separated the dynamics of the worm into two very different components: the $\{a_1,a_2\}$ plane with its  phase dynamics, responsible for crawling motions, and the mode $a_3$ which is connected with large curvature turns.  Because we found these modes as eigenvectors of a covariance matrix, we are guaranteed that the instantaneous amplitudes of the modes are not linearly correlated, but this does not mean that the dynamics of the different motions are completely uncoupled.   We found the clearest indications of mode  coupling between the phase in the $\{a_1,a_2\}$ plane and the amplitude $a_3$ at later times, which is illustrated by the correlation function in Fig \ref{fig:responses}b.  Intuitively, these correlations make sense, since the direction in which the worm forms its $\Omega$ shape must depend on the phase of the wave passing along the worm's body at the moment that the large amplitude bending is initiated.  This pattern of correlations is perturbed strongly by thermal stimuli.

The fact that the correlations between phase and the turning mode are stimulus dependent implies that the response of the turning mode to thermal stimuli depends on the phase which the worm finds itself at the time of the stimulus.    Perhaps some of the apparent randomness of turning responses thus is related to the fact that repeated thermal stimuli catch the worm at different initial phases.  As a preliminary test of this idea,  we implemented our analysis online, allowing an estimate of the phase with a delay of less than $125\,{\rm ms}$.  We then deliver an infrared pulse when the phase falls within a phase window that corresponds to either dorsal-- or ventral-- directed head swings.  To understand the predicted consequences of these pulses, consider an idealized example in which the target phase is $\pi/2$;  then the correlation structure in Fig \ref{fig:responses}b predicts a negative bias in $a_3$ which should generate right hand turns.  Similarly, if the target phase is $-\pi/2$, then continued thermal stimulation should generate left hand turns.  Note that the thermal pulse itself does not have a handedness, so that if the pulses are not synchronized to the state of the worm there should be no systematic preference for right vs. left handed turns.
These predictions are confirmed in Fig \ref{fig:responses}c.

\section{Discussion}

Our central result is that {\em C. elegans} motor behavior is vastly simpler than it could be, even when the organism moves freely.  Conceptually similar results have been obtained for aspects of motor control in humans and other primates, where postures or trajectories 
of limbs, hands or eyes 
are confined to spaces of low dimensionality despite the potential for more complex motions \cite{davella+bizzi_98,santello+al_98,sanger_00,osborne+al_05}.
For {\em C. elegans} itself, recent quantitative work has focused on simplifying behavior by matching to a discrete set of template behaviors, such as forward and backward motion of the center of mass \cite{hoshi+shingai_06},  sinusoidal undulations of the body \cite{karbowski+al_06}, or $\Omega$ bends \cite{huang+al_06}.  Our results combine and generalize these ideas.  Motor behaviors are described by projection of the body shape onto a small set of templates (the eigenworms), but the strengths of these projections vary continuously.   The templates are sinuous, but not sinusoidal, because the fluctuations in posture are not homogeneous along the length of the worm.  Our description of shape is intrinsic to the worm and invariant to the center of mass position, but motion in shape space predicts the center of mass motion.  There are discrete behavioral states, but these emerge as attractors of the underlying dynamics.  Most importantly, our choice of four eigenworms is driven not by hypotheses about the relevant components of behavior, but by the data itself.

The construction of the eigenworms guarantees that the instantaneous amplitudes along the different dimensions of shape space are not correlated linearly, but the dynamics of the different amplitudes are nonlinear and coupled; what we think of as a single motor action always involves coordinating multiple degrees of freedom.  Thus,  forward and backward motion correspond to positive and negative phase velocity in Fig \ref{fig:a1a2}, but transitions between these behavioral states occur preferentially at particular phases (data not shown).  Similarly, $\Omega$ turns involve a large amplitude excursion along $a_3$, but motion along this mode is correlated with phase in the ($\{a_1 , a_2\}$) plane, and this  correlation itself has structure in time (Fig \ref{fig:responses}b).  The problems of {\em C. elegans} motor control are simpler than for higher animals, but these nonlinear, coupled dynamics give a glimpse of the more general case.

Perhaps because of the strong coupling between the turning mode $a_3$ and the wriggling modes $a_1, a_2$, we have not found an equation of motion for $a_3$ alone which would be analogous to Eq's (\ref{dyn2}) for the phase.  Further work is required to construct a fully three dimensional dynamics which could predict the more complex correlations such as those in Fig \ref{fig:responses}b.    Turning should emerge from these equations not as another attractor, but as an `excitable' orbit analogous to the action potential in the Hodgkin--Huxley equations or to recent ideas about transient differentiation in genetic circuits \cite{suel+al_06}.  A major challenge would be to show that the stochastic dynamics of these equations can generate longer sequences of stereotyped events, such as pirouettes \cite{lockery99}.

We have shown that a meaningful set of behavioral coordinates can uncover deterministic responses. A response might seem stochastic or noisy because it depends on one or more behavioral variables that are not being considered.
In our experiments, nonlinear correlations among the behavioral variables suggest that some of the randomness in behavioral responses could be removed if sensory stimuli are delivered only when the worm is at a well defined initial state, and we confirmed this prediction by showing that phase--aligned thermal stimuli can `steer' the worm into trajectories with a definite chirality.  A crucial aspect of these experiments is that the stimulus is scalar---a temperature change in time has no spatial direction or handedness---but the response, by virtue of the correlation between stimulus and body shape, does have a definite spatial structure.
The alignment of thermal stimuli with the phase of the worm's movement in these experiments mimics the correlation between body shape and sensory input that occurs as the worm crawls in a thermal gradient, so the enhanced determinism of responses under these conditions may be connected to the computations which generate nearly deterministic isothermal tracking  \cite{Ryu02,Samuel06}.    

More generally, all behavioral responses have some mixture of deterministic and stochastic components.  In humans and other primates, it seems straightforward to create conditions that result in highly reproducible, stereotyped behaviors, such as reaching movements \cite{jeannerod88}.  At the opposite extreme, bacterial motility is modulated in response to sensory inputs, but these responses seem fundamentally probabilistic \cite{berg_book}.    Some of these differences may result from the physical nature of sensory stimuli in organisms of vastly different size \cite{berg+purcell_77,bialek_87}, but some of the differences may also result from differences of strategy or available computational power.  The more stochastic the response, the more challenging it is to characterize behavior quantitatively and to link behavior with underlying molecular and neural components, as is clear from recent work on {\em Drosophila} olfaction (see, for example, Ref \cite{voshall07}).  We hope that our approach to the analysis of behavior may help to uncover more deterministic components of the sensory--motor responses in other model organisms.
 
More than forty years of work on {\em C. elegans} has led to a  fully sequenced genome \cite{consortium_98} and to the complete wiring diagram of the nervous system \cite{white+al_86}.  Significant steps have been made toward the original dream  \cite{Brenner74} of connecting genes, neurons, and behavior \cite{bargmann_93,debono+maricq_05, lockery06}. Nonetheless, with the advances in molecular, cellular, and circuit  analyses, our ability to probe the mechanisms which generate behavior substantially exceeds our ability to characterize the behavior itself.    Perhaps our work  provides a step toward addressing this imbalance.

\section{Methods}
\textbf{Tracking microscopy.}
The imaging system consists of a Basler firewire CMOS camera (A601f, Basler, Ahrensburg, Germany) with 4x lens (55-901, Edmund Optics, Barrington, NJ) and a fiber optic trans-illuminator (DC-950, Dolan-Jenner, Boxborough, MA) mounted to an optical rail (Thorlabs, Newton, NJ). The rail is attached to a XY translation stage (Deltron, Bethel, CT) which is driven by stepper motors (US Digital, Vancouver, Washington). The stage driver is a home made unit utilizing a SimpleStep board (SimpleStep, Newton, NJ) and Gecko stepper motor drivers (Geckodrive, Santa Ana, CA). Image acquisition, processing, and stage driver control was done using LabVIEW (National Instruments, Austin, TX). Images of worms were isolated and identified using the image particle filter. A raw unprocessed JPEG image and a filtered process binary PNG image were written to the hard drive at rates up to 32 Hz. Concurrently at 4 Hz, the center of mass of the worm was calculated and the distance from the center of the field of view in pixels was computed. An error signal was then calculated via a coordinate transformation between the camera reference frame and the translational stage reference frame and the XY stage was moved to center the worm in the field of view.

\textbf{Worm preparation.}
The \textit{C. elegans} strain, N2, was grown at 20$^\circ$C and maintained under standard conditions \cite{Brenner74b}. Before each experiment, excess moisture from NGM assay plates (1.7\% Bacto Agar, 0.25\% Bacto-Peptone, 0.3\% NaCl, 1 mM CaCl$_2$, 1 mM MgSO$_4$, 25 mM potassium phosphate buffer, 5 $\mu$g/mL cholesterol) was removed by leaving them partially uncovered for 1 hr. A copper ring (5.1-cm inner diameter) pressed into the agar surface prevented worms from crawling to the side of the plate. Young adults were rinsed of \textit{E. coli} by transferring them with a worm pick from OP50 bacterial food plates into NGM buffer (same inorganic ion concentration as NGM assay plates) and letting them swim for 1 minute. Worms were transferred from the NGM buffer to the center of the assay plate (9-cm Petri dish). The plates were covered and tracking began after 1 minute and lasted no longer than 60 minutes. In the rare cases where worms stopped moving before the completion of the run, the data were excluded.

\textbf {Eigenworms.}  Images of worms captured by the worm tracker were processed using MATLAB (Mathworks, Natick, MA). Cases of self-intersection were excluded from processing. Images of worms were thinned to a single-pixel-thick backbone. A spline was fit through these points and then discretized into 101 segments, evenly spaced in units of the backbone arclength.  The $N=100$ angles between these segments were calculated and an overall rotation mode was removed by subtracting $\sum \theta(s(i))/N$ from each angle.  The shape covariance matrix $C(s,s')=\langle \left (\theta(s)-\langle\theta\rangle \right) \left( \theta(s^\prime)-\langle\theta\rangle \right ) \rangle$ was constructed from 9 freely crawling worms, sampled at 4\,Hz, for a period of 30 minutes.  Each eigenworm $u_{\mu}(s)$ is an eigenvector of the covariance matrix 
$\sum_{s^{\prime}} C(s,s^{\prime}) u_{\mu}(s^{\prime})=\lambda_\mu u_{\mu}(s)$.  The fractional variance captured by
$K$ eigenvectors is thus $\sigma^2_K=\sum_{\mu=1}^K \lambda_\mu/{\sigma^2}$, where $\sigma^2=\sum_\mu \lambda_\mu$ is the total variance of the measurements.   The same eigenworms shown in Fig.~\ref{fig:modes} were used throughout the various analysis reported in the paper.  The worm's phase was defined as $\phi=\tan^{-1}{\left(-a_2/ a_1\right)}$ where $a_1$ and $a_2$ were both normalized to unit variance.  

\textbf{Equations of motion.}
For the analysis of phase dynamics we sampled the worm shape at 32\,Hz.  Data for the construction of the equations of motion came from 12 worms, 5 trials per worm, with 4000 frames per trial. We also filtered each mode time series through a low-pass polynomial filter so that for each frame $(26\leq m \leq 3974)$, $\tilde{a}(m)=\sum_{n=-25}^{25} \sum_{j=0}^{4} p_j(m-n)^j$ where $\{p_j\}$ are the best-fit polynomial coefficients. Mode time derivatives were calculated using derivatives of the polynomial filter.  None of our  results depend critically on the properties of the filter.  The Langevin equations governing the phase dynamics are shown Eq.~(\ref{dyn2}) and we learn the functions $\{F(\phi,\omega),\sigma(\phi,\omega) \}$ directly from the time series \cite{langevinEOM1,langevinEOM2}.  By construction $\langle  \sigma[\phi(t),\omega(t)] \eta(t)  \rangle=0$ and therefore the optimal rms estimate of $F(\phi,\omega)$ is the conditional mean $\langle \dot{\omega}|\omega, \phi \rangle$.  We estimate $F$ by assuming a functional
expansion $F(\omega, \phi)=\sum_{m=-5}^{5} \sum_{p=0}^{5} \alpha_m^p w^k e^{-im\phi}$, where  the model parameters $\{\alpha_m^p \}$ were determined  by minimizing the rms error 
$\epsilon^2=\sum_t \left (\dot{\omega}(t) -F[\omega(t),\phi(t)]  \right)^2$ on training data ($90\%$) and the hyperparameters $\{m_{\rm max}=5,p_{\rm max}=5\}$  were chosen to minimize error on held-out data ($10 \%)$.   Once $F$ is known we can determine the noise in the system; we normalize $\langle\eta^2\rangle$ so that  $\sigma^2(\omega,\phi)=\langle \left(\dot{\omega}-F(\omega,\phi)\right)^2 |\omega,\phi \rangle$. 
The attractors contained within our derived dynamics were obtained by evolving initial
conditions spanning the sampled $\{\omega,\phi\}$ plane for long times (93.75\,s $\sim47$ cycles).   In the deterministic dynamics all trajectories evolve to one of four asymptotic states and we observed no switching.  

\textbf{Thermal impulse response (experiment).}
Worms were prepared as described earlier.  A collimated beam with a 1/e diameter of 5.6 mm (standard stimulus) or 1.5 mm (painful) from a 1440 nm diode laser (FOL1404QQM, Fitel, Peachtree City, GA) was positioned to heat the area covering the worm. The diode laser was driven with a commercial power supply and controller (Thorlabs, Newton, NJ). Power and duration of the beam was controlled through software using LabVIEW. For each worm, 1000 seconds of data was collected in cycles of 50 seconds. 12.5 seconds into each cycle the laser was turned on for a duration of 75 ms at 150 mW (standard) or 250 ms at 100mW (painful). The temperature increase caused by the laser pulses was measured using a 0.075mm T-type thermocouple (coco-003, Omega, Stamford, CT) placed on the surface of the agar and sampled with a thermocouple data acquisition device (USB-9211, National Instruments). For each measurement, 60 trials of 30 s cycles were averaged. The temperature increase was calculated by subtracting the maximum temperature (recorded immediately after the laser pulse) from the baseline temperature (recorded 9 s after the laser pulse). The temperature increase for the ÒstandardÓ pulse was 0.12$^\circ$C and the increase of the ÒpainfulÓ pulse was 0.73$^\circ$C.  

\textbf{Thermal impulse response (analysis).}
In Fig 6a, the time-dependent probability density $\rho_t(\omega)$  was smoothed before the onset of the impulse with a gaussian low-pass filter of size 0.19\,s in the $t$ direction and 0.17\,cycles/s in the
 $\omega$ direction.  In Fig 6b the correlation function $C(t,t^\prime)=\langle \left (\sin\phi(t)-\langle \sin\phi\rangle \right) \left (a_3(t^\prime)-\langle a_3 \rangle\right ) \rangle$ was calculated as follows.  Far from the time of the impulse (frames 800 to 1574, impulse on frame 400), we expect time-translation invariance $C_{\rm post}(t,t^\prime)=g(t-t^\prime) + \xi_{\rm post}(t,t^\prime)$ where $g(\Delta)=\langle C(i,j) \rangle_{i-j=\Delta}$ is the true correlation function and $\xi_{\rm post}(t,t^\prime)$ characterizes statistical error.  Similarly in a time window around the impulse (frames 24 to 800), $C_{\rm stim}(t,t^\prime)=g(t-t^\prime) + \xi_{\rm stim}(t,t^\prime)$.  However, the thermal impulse breaks this invariance and $\xi_{\rm stim}(t,t^\prime)$ contains both sampling fluctuations and stimulus-dependent correlation dynamics.  To separate these effects  we use singular value decomposition to compare $\xi_{\rm post}$ and $\xi_{\rm stim}$.
We write each matrix  $\xi_{\rm post/stim}(t,t^\prime)=\sum_{t^{\prime\prime}} U_{\rm post/stim}(t,t^{\prime\prime})S_{\rm post/stim}(t^{\prime\prime},t^{\prime\prime})V_{\rm post/stim}(t^{\prime\prime},t^\prime)$ and find that only two singular values of $\xi_{\rm stim}$ are significantly larger then $\xi_{\rm post}$. We then reconstruct the two-point function
 around the stimulus as $\tilde{C}_{stim}(t,t^\prime)=g(t-t^\prime)+\sum_{t^{\prime\prime}=1}^2 U_{\rm stim}(t,t^{\prime\prime})S_{\rm  stim}(t^{\prime\prime},t^{\prime\prime})V_{\rm stim}(t^{\prime\prime},t^\prime)$.

\textbf{Thermal steering.}
Preparation of worms and instrumentation were the same as described for the thermal impulse response. However, instead of processing worm images off-line, real-time calculation of the eigenworms and shape phase $\phi$ was done using custom dynamic-linked image processing libraries written in C along with supporting LabVIEW code. The modes were computed as previously described except that the spline interpolation algorithm was replaced with a Hermitian interpolation algorithm to reduce the processing time. The processing time was short enough to simultaneously track and calculate modes at 8\,Hz. For phase dependent measurements, the laser was fired when the worm was moving forward and $\phi$ fell within a prescribed interval (width 1 radian). The laser pulse (150 mW) lasted for 75 ms and caused a temperature increase of 0.12$^\circ$ C. For each run a pair of triggering phase windows (0 to -1, and 2.1 and 3.1 radians) corresponding to the dorsal- and ventral-directed head swing was used. The sequence of each run started with a 5 minute period of no stimulus followed by the pair of phase dependent stimuli. The order of each pair of stimulus conditions was switched for each successive run. For the randomized pulse control experiments, the laser was fired with a uniform phase probability, but with conditions that restricted the firing interval to be longer than 2 seconds.

\textbf{Steering and turn identification.} 
The time-average change in orientation of the wormÕs path, $\langle\dot \Theta \rangle$  (rad/s), was calculated from the angular changes between the positions of the center of mass of the worm during forward runs of at least 4\,s in length. Given positions ($r_1, r_2, r_3, \ldots, r_N$), the angles between connecting segments ($r_2-r_1, r_3-r_2, \ldots , r_N- r_{(N-1)}$) were calculated. 
$\langle\dot \Theta \rangle$ was calculated in intervals of 10\,s. Since the distributions were Gaussian (data not shown) with similar variance, we used the Student's t-test to determine if the values of $\dot\Theta$  under thermal stimulation were significantly different than the control ($p< 0.05$).  Since we were interested in the change in orientation during forward motion we excluded trajectory data that contained large turns or reversals along with angular changes greater than $\pi/4$ radians.  These events were
automatically detected by measuring the compactness of the worm shape.  Compactness was calculated by measuring the longest distance between two points in the worm shape (also known as the max feret distance) and normalizing this with the maximum value for the entire data run. Turns were flagged when the compactness fell below 0.6. 

\bibliographystyle{abbrv}

\begin{thebibliography}{99}
\bibitem{green+swets_66}
DM Green \& JA Swets, {\em Signal Detection Theory and Psychophysics}
(John Wiley \& Sons, New York 1966).
\bibitem{ethology}
J Bolhuis \& L Giraldeau, Editors, {\em The Behavior of Animals: Mechanism, Function and Evolution}  (Blackwell Publishing, Oxford, 2004).
\bibitem{geng+al_03}
W Geng, P Cosman, JH  Baek, C Berry \& WR Schafer, Quantitative classification and natural clustering of {\em Caenorhabditis elegans} behavioral phenotypes. {\em Genetics} {\bf 165,} 1117--1126 (2003).
\bibitem{cronin+al_05}
CJ Cronin, JE Mendel , S Mukhtar, Y Kim, RC Stirbl , J Bruck  \& PW Sternberg, An automated system for measuring parameters of nematode sinusoidal movement. {\em BMC Genet} {\bf 6,} 5--24 (2005).
\bibitem{hoshi+shingai_06}
K Hoshi \& R  Shingai, Computer--driven automatic identification of locomotion states in {\em Caenorhabditis elegans}. {\em J Neurosci Meth} {\bf 157,} 355--363 (2006).
\bibitem{karbowski+al_06}
J Karbowski, CJ Cronin, A Seah, JE Mendel, D Cleary \& PW  Sternberg, Conservation rules, their breakdown, and optimality in {\em Caenorhabditis} sinusoidal locomotion. {\em J Theor Biol} 
{\bf 242,} 652--669 (2006).
\bibitem{huang+al_06}
KM Huang, PC Cosman \& W Schafer, Machine vision based detection of omega bends and reversals in {\em C. elegans}. {\em J Neurosci Meth} {\bf  158,} 323--336 (2006).
\bibitem{croll_75} N Croll, Components and patterns in the behavior of the nematode {\it Caenorhabditis elegans}. {\em J Zool}  {\bf 176,} 159--176 (1975).
\bibitem{bargmann_93}
CI Bargmann,   Genetic and cellular analysis of behavior in {\em C. elegans}. {\em  Annu Rev Neurosci} {\bf  16,} 47--71 (1993).
\bibitem{hedgecock+russell_75}
EM Hedgecock \& RL Russell, Normal and mutant thermotaxis in the nematode {\em Caenorhabditis elegans}. {\em Proc Nat'l Acad Sci (USA)} {\bf 72,} 4061--4065 (1975).
\bibitem{chalfie+al_85}
M Chalfie, JE Sulston, JG White, E Southgate, JN Thomson \& S Brenner, The neural circuit for touch sensitivity in {\em Caenorhabditis elegans}. {\em J Neurosci} {\bf 5,}  956--964 (1985).
\bibitem{frenet}
J  Frenet, Sur  quelque properiet\`{e}s des courbes \'{a} double courbure. Th\`{e}se, Toulouse, 1847.
Abstract in {\em Jour de Math} {\bf 17,} 437 (1852).
\bibitem{struik_61}
DJ Struik, {\em Lectures on Classical Differential Geometry} (Addison-Wesley, Reading, 1961).
\bibitem{pain} 
N Wittenburg \& R Baumeister, Thermal avoidance in {\em Caenorhabditis elegans}: An approach to the study of nociception.
{\em Proc Nat'l Acad Sci (USA)} {\bf 96,} 10477--10482 (1999).
\bibitem{stat_mech_text}
NG van Kampen, {\em Stochastic Processes in Physics and Chemistry, 2nd Edition} (North Holland, Amsterdam, 2001).
\bibitem{flyvbjerg}
D Selmecai, S Mosler, PH Hagedorn, NB Larsen \& H Flyvbjerg,
Cell motility as persistent random motion:  Theories from experiment.
{\em Biophys J} {\bf 89,} 912--931 (2005).
\bibitem{gray+al_05} 
JM Gray, JJ Hill \& CI  Bargmann, A circuit for navigation in {\em Caenorhabditis elegans}. {\em Proc Nat'l Acad Sci (USA)}  {\bf 102,} 3184--3191 (2005).
\bibitem{Lockery03} 
HA Zariwala, AC Miller, S Faumont \& SR Lockery, Step response analysis of thermotaxis in {\em Caenorhabditis elegans}.  {\em J Neurosci} {\bf 23,} 4369--4377 (2003).
\bibitem{Ryu02} 
WS Ryu \& AD Samuel, Thermotaxis in {\em Caenorhabditis elegans} analyzed by measuring responses to defined thermal stimuli. {\em J Neurosci} {\bf 22,} 5727--5733 (2002).
\bibitem{davella+bizzi_98}
A d'Avella \& E Bizzi, Low dimensionality of surpaspinally induced force fields. {\em Proc Nat'l Acad Sci (USA)} {\bf  95,} 7711--7714 (1998).
\bibitem{santello+al_98}
M Santello, M Flanders \& JF Soechting, Postural hand strategies for tool use. {\em  J Neurosci} {\bf 18,} 10105--10115 (1998).
\bibitem{sanger_00}
TD Sanger, Human arm movements described by a low-dimensional superposition of principal components. {\em J Neurosci} {\bf 20,} 1066--1072 (2000).
\bibitem{osborne+al_05}
LC Osborne, SG Lisberger \& W Bialek, A sensory source for motor variation.
{\em Nature} {\bf 437,} 412--416 (2005).
\bibitem{suel+al_06}
GM Suel, J Garcia--Ojalvo, LM Liberman \& MB Elowitz, An excitable gene regulatory circuit induces transient cellular differentiation. {\em Nature} {\bf 440,} 545--550 (2006).
\bibitem{lockery99} 
JT Pierce--Shimomura, TM Morse \& SR  Lockery, The fundamental role of pirouettes in {\it Caenorhabditis elegans} chemotaxis. {\em J Neurosci} {\bf 19,} 9557--9569 (1999).
\bibitem{Samuel06}
L Luo, DA Clark, D Biron, L Mahadevan \& AD Samuel, Sensorimotor control during isothermal tracking in Caenorhabditis elegans.  {\em J Exp Biol} {\bf 209,} 4652--4662 (2006).
\bibitem{jeannerod88}
M Jeannerod, {\em The Neural and Behavioural Organization of Goal-Directed Movements} (Oxford Univ Press, Oxford, 1988).
\bibitem{berg_book}
HC Berg, {\em E. coli in Motion} (Springer--Verlag, New York, 2004).
\bibitem{berg+purcell_77}
HC Berg \& EM Purcell, Physics of chemoreception.  {\em Biophys J} {\bf 20,}  
193--219 (1977).
\bibitem{bialek_87}
W Bialek, Physical limits to sensation and perception.  {\em Annu Rev Biophys Biophys Chem} {\bf  16,}  455--478 (1987).
\bibitem{voshall07}
A Keller \& LB Vosshall, lnfluence of odorant receptor repertoire on odor perception in humans and fruit flies. {\em Proc Nat'l Acad Sci (USA)} {\bf 104,} 5614--5619 (2007).
\bibitem{consortium_98}
The {\em C. elegans} Sequencing Consortium, Genome sequence of the nematode {\em C. elegans}: a platform for investigating biology. {\em Science} {\bf 282,} 2012--2018 (1998).
\bibitem{white+al_86}
JG White, E Southgate, JN Thomson \& S Brenner, The structure of the nervous system of the nematode {\em Caenorhabditis elegans}. {\em  Phil Trans R Soc B}  {\bf 314,} 1--340 (1986).
\bibitem{Brenner74}
S Brenner,  The genetics of {\it Caenorhabditis elegans}.
{\em Genetics} {\bf 77,} 71--94 (1974).
\bibitem{debono+maricq_05}
M de Bono \& AV Maricq, Neuronal substrates of complex behaviors in {\em C. elegans}. {\em Annu Rev Neurosci} {\bf 28,} 451--501 (2005).
\bibitem{lockery06} S Faumont \& SR Lockery, The awake behaving worm: simultaneous imaging of neuronal activity and behavior in intact animals at millimeter scale. {\em J Neurophys} {\bf 95,} 1976--1981  (2006).
\bibitem{Brenner74b}
JE Sulston \& S Brenner,  The DNA of {\it Caenorhabditis elegans}.
{\em Genetics} {\bf 77,} 95--104 (1974).
\bibitem{langevinEOM1} R Friedrich, J Peinke \& Ch Renner, How to quantify deterministic and random influences on the statistics of the foreign exchange market. {\em Phys Rev Lett} {\bf 84,} 5224--5227 (2000).
\bibitem{langevinEOM2} E Racca \& A Porporato, Langevin equations from time series. {\em Phys Rev E} {\bf 71,} 027101--027103 (1999).
\end{thebibliography}

{\small
We thank D Chigirev, SE Palmer, E Schneidman and G Tka\v{c}ik for discussions and AR Chapman for help in the initial building of the worm tracker and for programming the thinning algorithm used for real--time processing. This work was supported in part by Los Alamos National Laboratory, by a training grant from the National Institute of Mental Health, by NIH Grant P50 GM071508 and EY017210, and by the Swartz Foundation.  WB also thanks LF Abbott, KD Miller, and the Center for Theoretical Neuroscience at Columbia University for their hospitality.}  
\end{document}